# 1

# VLSI Architectures for WIMAX Channel Decoders

Maurizio Martina and Guido Masera
*Politecnico di Torino*
*Italy*

## 1. Introduction

WIMAX has gained a wide popularity due to the growing interest and diffusion of broadband wireless access systems. In order to be flexible and reliable WIMAX adopts several different channel codes, namely convolutional-codes (CC), convolutional-turbo-codes (CTC), block-turbo-codes (BTC) and low-density-parity-check (LDPC) codes, that are able to cope with different channel conditions and application needs.

On the other hand, high performance digital CMOS technologies have reached such a development that very complex algorithms can be implemented in low cost chips. Moreover, embedded processors, digital signal processors, programmable devices, as FPGAs, application specific instruction-set processors and VLSI technologies have come to the point where the computing power and the memory required to execute several real time applications can be incorporated even in cheap portable devices.

Among the several application fields that have been strongly reinforced by this technology progress, channel decoding is one of the most significant and interesting ones. In fact, it is known that the design of efficient architectures to implement such channel decoders is a hard task, hardened by the high throughput required by WIMAX systems, which is up to about 75 Mb/s per channel. In particular, CTC and LDPC codes, whose decoding algorithms are iterative, are still a major topic of interest in the scientific literature and the design of efficient architectures is still fostering several research efforts both in industry and academy.

In this Chapter, the design of VLSI architectures for WIMAX channel decoders will be analyzed with emphasis on three main aspects: performance, complexity and flexibility. The chapter will be divided into two main parts; the first part will deal with the impact of system requirements on the decoder design with emphasis on memory requirements, the structure of the key components of the decoders and the need for parallel architectures. To that purpose a quantitative approach will be adopted to derive from system specifications key architectural choices; most important architectures available in the literature will be also described and compared.

The second part will concentrate on a significant case of study: the design of a complete CTC decoder architecture for WIMAX, including also hardware units for depuncturing (bit-deselection) and external deinterleaving (sub-block deinterleaver) functions.

## 2. From system specifications to architectural choices

The system specifications and in particular the requirement of a peak throughput of about 75 Mb/s per channel imposed by the WIMAX standard have a significant impact on the decoder architecture. In the following sections we analyze the most significant architectures proposed in the literature to implement CC decoders (Viterbi decoders), BTC, CTC and LDPC decoders.

### 2.1 Viterbi decoders

The most widely used algorithm to decode CCs is the Viterbi algorithm [Viterbi, 1967], which is based on finding the shortest path along a graph that represents the CC trellis. As an example in Fig. 1 a binary 4-states CC is shown as a feedback shift register (a) together with the corresponding state diagram (b) and trellis (c) representations.

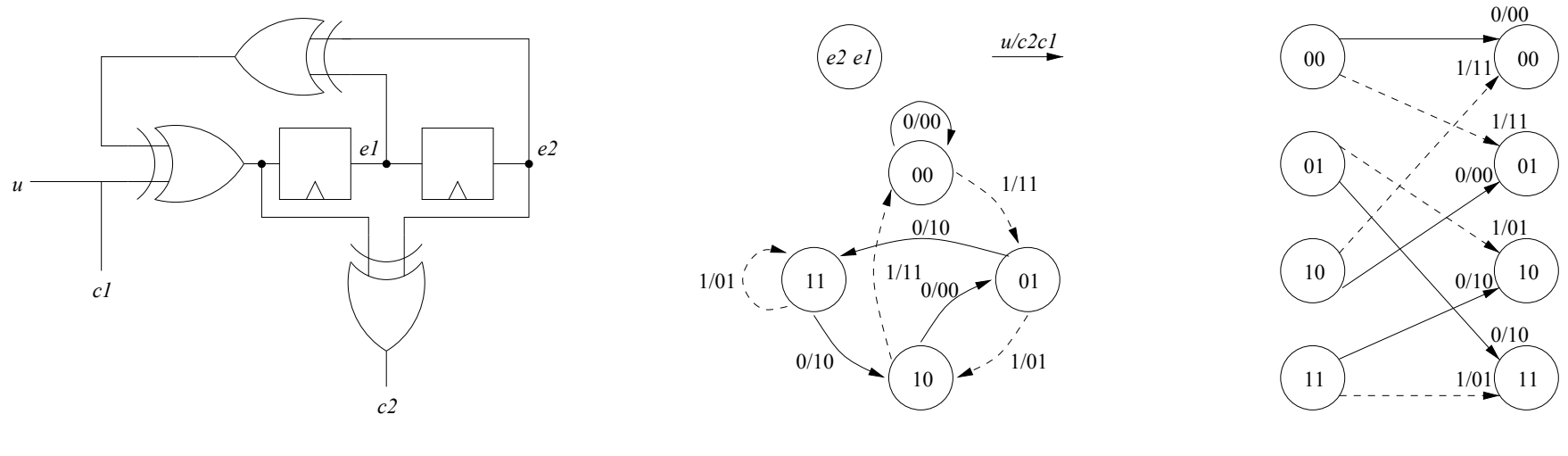

Fig. 1. Binary 4-state CC example: shift register (a), state diagram (b) and trellis (c) representations

In the given example, the feedback shift register implementation of the encoder generates two output bits, $c_1$ and $c_2$ for each received information bit, $u$; $c_1$ is the systematic bit. The state diagram basically is a Mealy finite state machine describing the encoder behaviour in a time independent way: each node corresponds to a valid encoder state, represented by means of the flip flop content, $e_1$ and $e_2$, while edges are labelled with input and output bits. The trellis representation also provides time information, explicitly showing the evolution from one state to another in different time steps (one single step is drawn in the picture).
At each trellis step $n$, the Viterbi algorithm associates to each trellis state $S$ a state metric $\Gamma^S_n$ that is calculated along the shortest path and stores a decision $d^S_n$, which identifies the entering transition on the shortest path. First, the decoder computes the branch metrics ($\gamma_n$), that are the distances from the metrics labelling each edge on the trellis and the actual received soft symbols. In the case of a binary CC with rate 0.5 the soft symbols are $\lambda 1_n$ and $\lambda 2_n$ and the branch metrics $\gamma_n(c2,c1)$ (see Fig. 2 (a)). Starting from these values, the state metrics are updated by selecting the larger metric among the metrics related to each incoming edge of a trellis state and storing the corresponding decision $d^S_n$. Finally, decoded bits are obtained by means of a recursive procedure usually referred to as trace-back. In order to estimate the sequence of bits that were encoded for transmission, a state is first selected at the end of the trellis portion to be decoded, then the decoder iteratively goes backward through the state history memory where decisions $d^S_n$ have been previously stored: this allows one to select, for current state, a new state, which is listed in the state history trace as being the predecessor to that state. Different implementation methods are available to make the initial state choice and to size the portion of trellis where the trace back operation is performed: these methods affect both decoder complexity and error correcting capability. For further details on the algorithm the reader can refer to [Viterbi, 1967]; [Forney, 1973]. Looking at the global architecture, the main blocks required in a Viterbi decoder are the branch metric unit (BMU) devoted to compute $\gamma_n$, the state metric unit (SMU) to calculate $\Gamma^S_n$ and the trace-back unit (TBU) to obtain the decoded sequence. The BMU is made of adders and subtracters to properly combine the input soft symbols (see Fig. 2 (a)). The SMU is based on the so called add-compare select structure (ACS) as shown in Fig.2 (b). Said $i$ the $i$-th starting state that is connected to an arriving state $S$ by an edge whose branch metric is $\gamma^i_{n-1}$, then $\Gamma^S_n$ is calculated as in (1).

$$\Gamma^S{}_n = \max_i \{\Gamma^i{}_{n-1} + \gamma^i{}_{n-1}\} \quad (1)$$

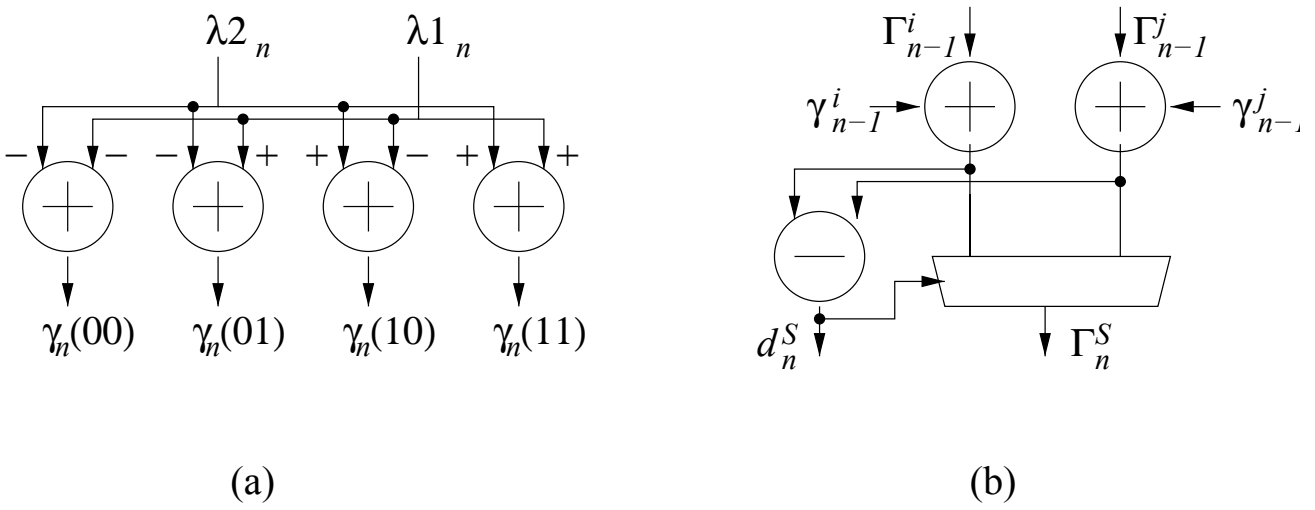

Fig. 2. BMU and ACS architectures for a rate 0.5 CC

As it can be inferred from (1) $\Gamma^S_n$ is obtained by adding branch metrics with state metrics, comparing and selecting the higher metric that represents the shortest incoming path. The corresponding decision $d^S_n$ is stored in a memory that is later read by the TBU to reconstruct the survived path. Due to the recursive form of (1), as long as $n$ increases, the number of bits to represent $\Gamma^S_n$ tends to become larger. This problem can be solved by normalizing the state metrics at each step. However, this solution requires to add a normalization stage increasing both the SMU complexity and critical path. An effective technique, based on two complement representation, helps limiting the growth of state metrics, as described in [Hekstra, 1989].

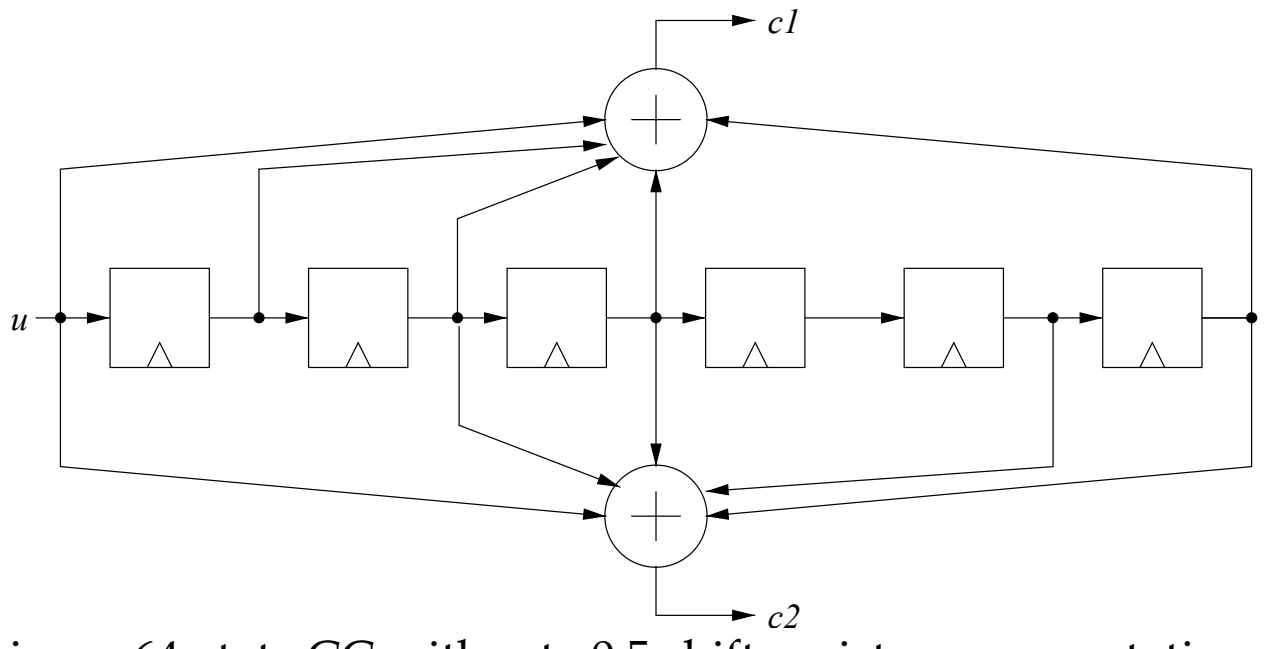

Fig. 3. WIMAX binary 64-state CC with rate 0.5 shift register representation

The WIMAX standard specifies a binary 64 states CC with rate 0.5, whose shift register representation is shown in Fig. 3. Usually Viterbi decoder architectures exploit the trellis intrinsic parallelism to simultaneously compute at each trellis step all the branch metrics and update all the state metrics. Thus, said $n$ the number of states of a CC, a parallel architecture employs a BMU and $n$ ACS modules. Moreover, to reduce the decoding latency, the trace-back is performed as a sliding-window process [Radar, 1981] on portions of trellis of width $W$. This approach not only reduces the latency, but also the size of the decision memory that depending on the TBU radix requires usually $3W$ or $4W$ cells [Black & Meng, 1992].
To improve the decoder throughput, two [Black & Meng, 1992] or more [Fettweis & Meyr, 1989]; [Kong & Parhi, 2004]; [Cheng & Parhi, 2008] trellis steps can be processed concurrently. These solutions lead to the so called higher radix or $M$-look-ahead step architectures. According to [Kong & Parhi, 2004], the throughput sustained by an $M$-look-ahead step architecture, defined as the number of decoded bits over the decoding time is

$$T = \frac{k \cdot N_T \cdot f_{clk}}{N_T / M + W} \approx f_{clk} \cdot M \cdot k \quad (2)$$

where $f_{clk}$ is the clock frequency, $N_T$ is the number of trellis steps, $k$=1 for a binary CC, $k$=2 for a double binary CC and the right most expression is obtained under the condition $W << N_T$ that is a reasonable assumption in real cases.
Thus, to achieve the throughput required by the WIMAX standard with a clock frequency limited to tens to few thousands of MHz, $M$=1 (radix-2) or $M$=2 (radix-4) is a reasonable choice.
However, since CCs are widely used in many communication systems, some recent works as [Batcha & Shameri, 2007] and [Kamuf et al., 2008] address the design of flexible Viterbi decoders that are able to support different CCs. As a further step [Vogt & When, 2008] proposed a multi-code decoder architecture, able to support both CCs and CTCs.

### 2.2 BTC decoders

Block Turbo Codes or product codes are serially concatenated block codes. Given two block codes $C_1=(n_1,k_1,\delta_1)$ and $C_2=(n_2,k_2,\delta_2)$ where $n_i$, $k_i$ and $\delta_i$ represent the code-word length, the number of information bits, and the minimum Hamming distance, respectively, the corresponding product code is obtained according to [Pyndiah, 1998] as an array with $k_1$ rows and $k_2$ columns containing the information bits. Then coding is performed on the $k_1$ rows with $C_2$ and on the $n_2$ obtained columns with $C_1$. The decoding of BTC codes can be performed iteratively row-wise and column-wise by using the sub-optimal algorithm detailed in [Pyndiah, 1998]. The basic idea relies on using the Chase search [Chase, 1972] a near-maximum-likelihood (near-ML) searching strategy to find a list of code-words and an ML decided code-word $\underline{d}=\{d_0,\ldots, d_{n-1}\}$ with $d_j \in \{-1,+1\}$. According to the notation used in [Vanstraceele et al., 2008], decision reliabilities are computed as

$$\lambda(d_j) \approx \frac{|\underline{r} - \underline{c}^{-1(j)}|^2 - |\underline{r} - \underline{c}^{+1(j)}|^2}{4} \quad (3)$$

where $\underline{r}=\{r_0,\ldots r_{n-1}\}$ is the received code-word and $c^{-1(j)}$ and $c^{+1(j)}$ are the code-words in the Chase list at minimum Euclidean distance from $\underline{r}$ such that the $j$-th bit of the code-word is -1 and +1 respectively. Then one decoder sends to the other the extrinsic information

$$w_j^{out} = \lambda(d_j) - r_j \tag{4}$$

If the Chase search fails the extrinsic information is approximated as

$$w_j^{out} = \beta \cdot d_j \tag{5}$$

where $\beta$ is a weight factor increasing with the number of iterations.
The decoder that receives the extrinsic information uses an updated version of $\underline{r}$ obtained as

$$r_j^{new} = r_j^{old} + \alpha \cdot w_j^{in} \tag{6}$$

where $\alpha$ is a weight factor increasing with the number of iterations. A scheme of the elementary block turbo decoder is shown in Fig. 4 where the block named "decoder" is a Soft-In-Soft-out (SISO) module that performs the Chase search and implements (3), (4) and (5). An effective solution to implement the SISO module is based on a three pipelined stage architecture where the three stages are identified as reception, processing, and transmission units [Kerouedan & Adde, 2000]. As detailed in [LeBidan et al., 2008], during each stage, the $N$ soft values of the received word $r$ are processed sequentially in $N$ clock periods. The reception stage is devoted to find the least reliable bits in the received code-word. The processing stage performs the Chase search and the transmission stage calculates $\lambda(d_j)$, $w_j$ and $r_j^{new}$. Another solution is proposed in [Goubier et al. 2008] where the elementary decoder is implemented as a pipeline resorting to the mini-maxi algorithm, namely by using mini-maxi arrays to store the best metrics of all decoded code-words in the Chase list.

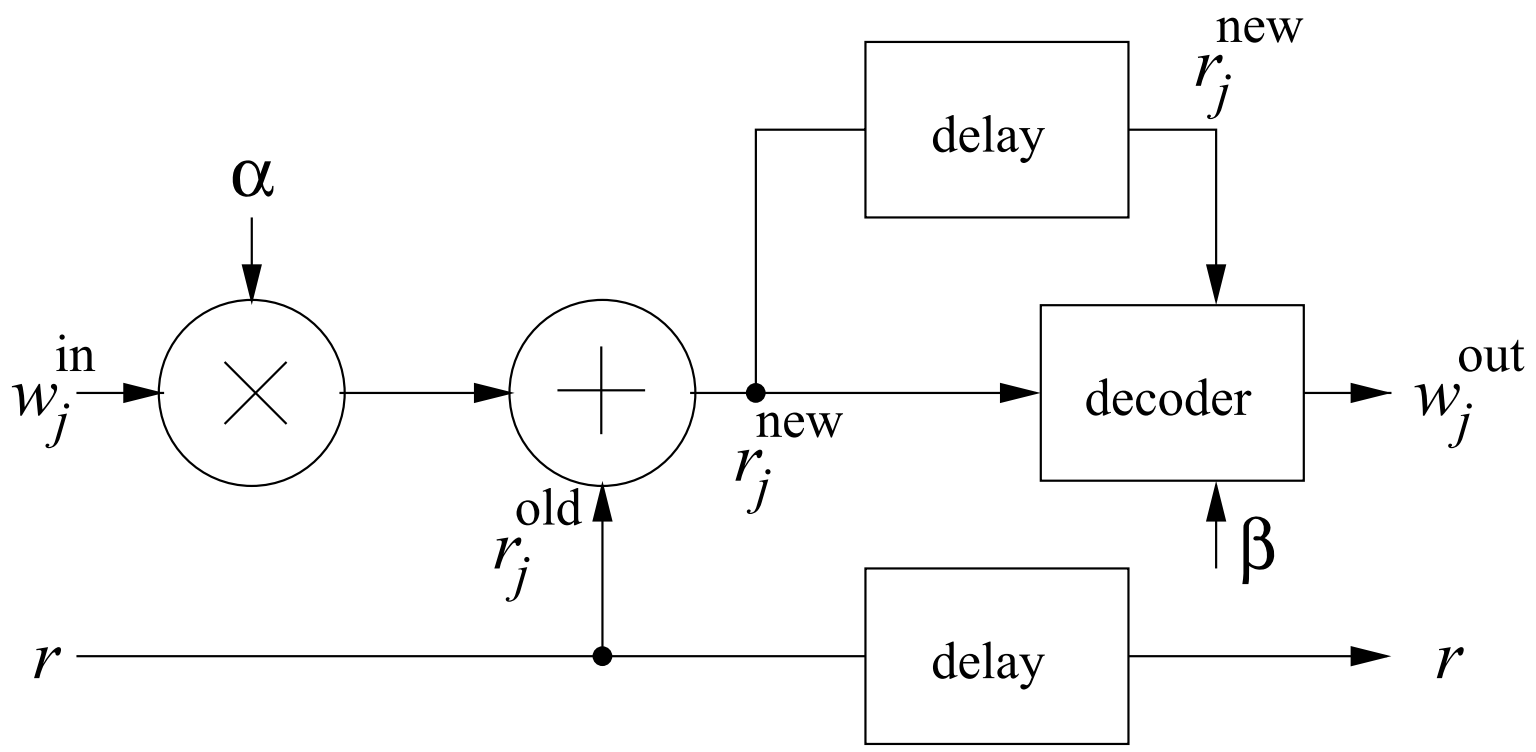

Fig. 4. Elementary block turbo decoder scheme

Several works in the literature deal with BTC complexity reduction. As an example [Adde & Pyndiah, 2000] suggests to compute $\beta$ in (5) on a per-code-word basis, whereas in [Chi et al., 2004] the dependency on $\alpha$ in (6) is solved by replacing the term $\alpha \cdot w_j$ with $\tanh(w_j/2)$. In [Le et al. 2005] both $\alpha$ in (6) and $\beta$ in (5) are avoided by exploiting Euclidean distance property.
Due to its row-column structure, the block turbo decoder can be parallelized by instantiating several elementary decoders to concurrently process more rows or columns, thus increasing the throughput. As a significant example in [Jego et al., 2006] a fully parallel BTC decoder is proposed. This solution instantiates $n_1+n_2$ decoders that work concurrently. Moreover, by properly managing the scheduling of the decoders and interconnecting them through an Omega network intermediate results (row decoded data or column decoded data) are not stored.
A detailed analysis of throughput and complexity of BTC decoder architectures can be found in [Goubier et al. 2008] and [LeBidan et al., 2008]. In particular, according to [Goubier et al. 2008] a simple one block decoder architecture that performs the row/column decoding sequentially (interleaved architecture) requires $2(n_1+n_2)$ cycles to complete an iteration; as a consequence it achieves a throughput

$$T = \frac{k_1 \cdot k_2 \cdot f_{clk}}{I \cdot 2(n_1 + n_2)} \tag{7}$$

where $I$ is the number of iterations and $f_{clk}$ is the clock frequency. The BTC specified for WIMAX is obtained using twice a binary extended Hamming code out of the ones show in Table 1

| *N* | *k* |
|---|---|
| 15 | 11 |
| 31 | 26 |
| 63 | 57 |

Table 1. WIMAX binary extended Hamming codes ($H(n,k)$) used for BTC

Considering the interleaved architecture described in [Goubier et al. 2008] where a fully decoded block is output every 4.5 half iterations, we obtain that 75 Mb/s can be obtained with a clock frequency of 84 MHz, 31 MHz and 14 MHz for $H(15,11)$, $H(31,26)$ and $H(63,57)$ respectively.

### 2.3 CTC decoders

Convolutional turbo codes were proposed in 1993 by Berrou, Glavieux and Thitimajshima [Berrou et al., 1993] as a coding scheme based on the parallel concatenation of two CCs by the means of an interleaver (Π) as shown in Fig. 5 (a). The decoding algorithm is iterative and is based on the BCJR algorithm [Bahl et al., 1974] applied on the trellis representation of each constituent CC (Fig. 5 (b)). The key idea relies on the fact that the extrinsic information output by one CC is used as an updated version of the input a-priori information by the other CC. As a consequence, each iteration is made of two half iterations, in one half iteration the data are processed according to the interleaver ($\Pi$) and in the other half iteration according to the deinterleaver ($\Pi^{-1}$). The same result can be obtained by implementing an in-order read/write half iteration and a scrambled (interleaved) read/write half iteration. The basic block in a turbo decoder is a SISO module that implements the BCJR algorithm in its logarithmic likelihood ratio (LLR) form. If we consider a Recursive Systematic CC (RSC code), the extrinsic information $\lambda_k(u;O)$ of an uncoded symbol $u$ at trellis step $k$ output by a SISO is

$$\lambda_k(u;O) = \max_{e:u(e)=u}{}^{*}\{b(e)\} - \max_{e:u(e)=\tilde{u}}{}^{*}\{b(e)\} - \lambda_k(u;I) - \pi_k(c^u;I) \tag{8}$$

where $\tilde{u}$ is an uncoded symbol taken as a reference (usually $\tilde{u}=\mathbf{0}$), $e$ represents a certain transition on the trellis and $u(e)$ is the uncoded symbol $u$ associated to $e$. The max* function is usually implemented as a max followed by a correction term [Robertson et al., 1995]; [Gross & Gulak, 1998]; [Cheng & Ottosson, 2000]; [Classon et al., 2002]; [Wang et al., 2006]; [Talakoub et al. 2007]. A scaling factor can also be applied to further improve the max or max* approximation [Vogt & Finger, 2000]. The correction term, usually adopted when decoding binary codes, can be omitted for double binary turbo codes [Berrou et al. 2001] with minor error rate performance degradation. The term $b(e)$ in (8) is defined as

$$b(e) = \alpha_{k-1}[s^S(e)] + \gamma_k[e] + \beta_k[s^E(e)] \tag{9}$$

$$\alpha_k[s] = \max_{e:s^E(e)=s}\{\alpha_{k-1}[s^S(e)] + \gamma_k[e]\} \tag{10}$$

$$\beta_k[s] = \max_{e:s^S(e)=s}\{\beta_{k+1}[s^E(e)] + \gamma_k[e]\} \tag{11}$$

$$\gamma_k[e] = \pi_k[u(e);I] + \pi_k[c(e);I] \tag{12}$$

where $s^S(e)$ and $s^E(e)$ are the starting and the ending states of $e$, $\alpha_k[s^S(e)]$ and $\beta_k[s^E(e)]$ are the forward and backward state metrics associated to $s^S(e)$ and $s^E(e)$ respectively (see Fig. 5 (b)) and $\gamma_k[e]$ is the branch metric associated to $e$. The $\pi_k[c(e);I]$ term is computed as a weighted sum of the $\lambda_k[c;I]$ produced by the soft demodulator as

$$\pi_k[c(e);I] = \sum_{i}^{n_c} c_i(e)\lambda_k[c_i(e);I] \tag{13}$$

where $c_i(e)$ is one of the coded bits associated to $e$ and $n_c$ is the number of bits forming a coded symbol $c$ and $\pi_k[c^u(e);I]$ in (8) is obtained as $\pi_k[c(e);I]$ considering only the systematic bits corresponding to the uncoded symbol $u$ out of the $n_c$ coded bits. The $\pi_k[u(e);I]$ term is obtained combining the input a-priori information $\lambda_k(u;I)$ and for a double binary code can be written as in (14), where $A$ and $B$ represent the two bits forming an uncoded symbol $u$.
The CTC specified in the WIMAX standard is based on a double binary 8-state constituent CC as shown in Fig. 6, where each CC receives two uncoded bits ($A$, $B$) and produces four coded bits, two systematic bits ($A$,$B$) and two parity bits ($Y$,$W$). As a consequence, at each trellis step four transitions connect a starting state to four possible ending states. Due to the trellis symmetry only 16 branch metrics out of the possible 32 branch metrics are required at each trellis step. As pointed out in [Muller et al. 2006] high throughput can be achieved by exploiting the trellis parallelism, namely computing concurrently all the branch and state metrics.

$$\pi_k(u;I) = \begin{cases} 0 & if \quad u(e) = (A='0', B='0') \\ \lambda_k^{\bar{A}B} & if \quad u(e) = (A='0', B='1') \\ \lambda_k^{A\bar{B}} & if \quad u(e) = (A='1', B='0') \\ \lambda_k^{AB} & if \quad u(e) = (A='1', B='1') \end{cases} \tag{14}$$

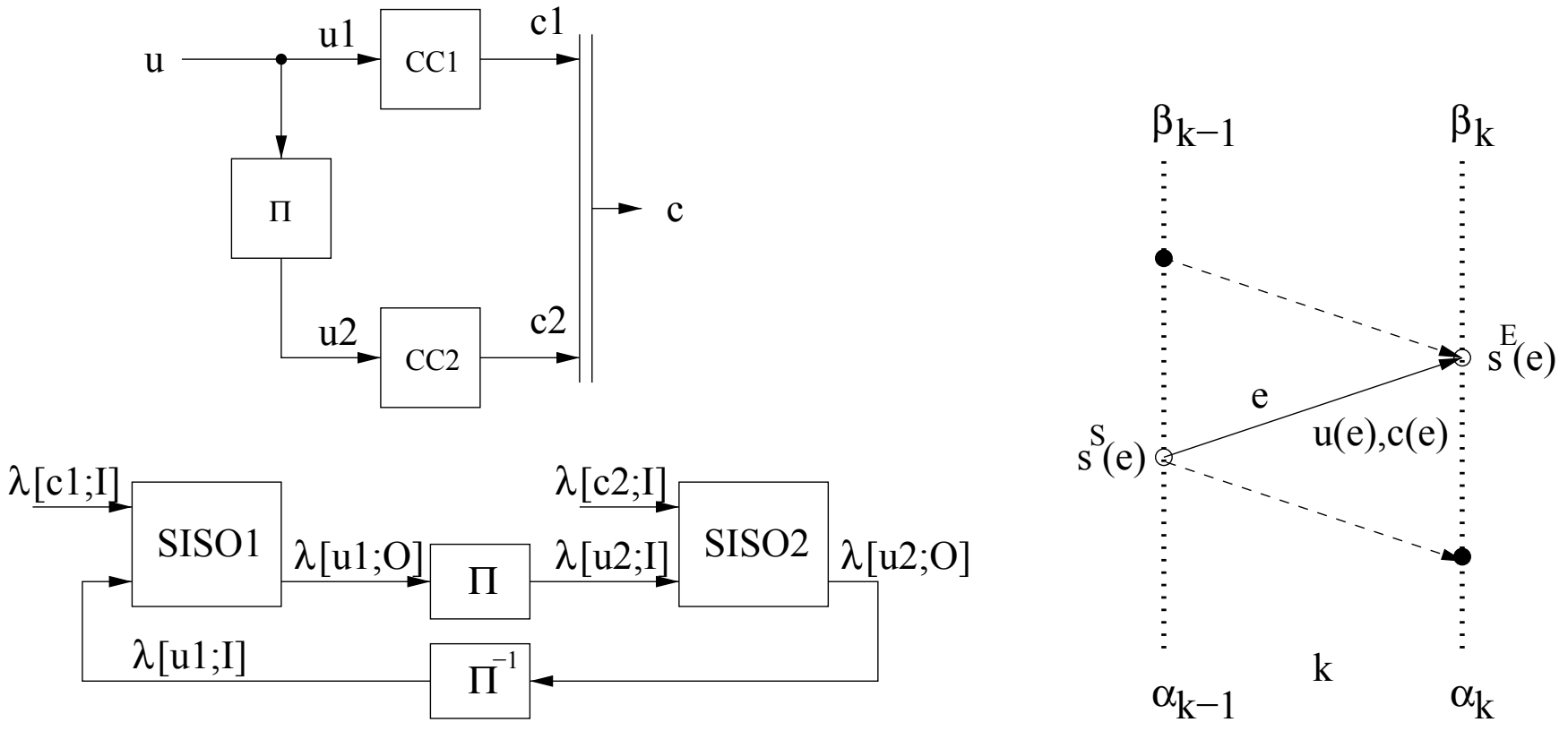

(a) (b)

Fig. 5. Convolutional turbo code: coder and iterative SISO based decoder (a), notation for a trellis step in the SISO (b)

The 16 branch metrics are computed by a BMU that implements (12) as shown in Fig. 7. To reduce the latency of the SISO, usually the decoding is based on a sliding-window approach [Benedetto et al., 1996]. As a consequence, at least two BMUs are required to compute the two recursions (forward and backward) according to the BCJR algorithm. However, since β metrics require to be trained between successive windows, usually a further BMU is required. A solution based on the inheritance of the border metrics of each window [Abbasfar & Yao 2003] requires only two BMUs. Furthermore, this strategy reduces the SISO latency to the sliding window width $W$. The state metrics are updated according to (10) and (11) by two state metric processors, each of which is made of a proper number of processing elements (PE). As shown in Fig. 7 for the WIMAX CTC 8 PEs are required. It is worth pointing out that the constituent codes of the WIMAX CTC use the circulation state tailbiting strategy proposed in [Weiss et al. 2001] that ensures that the ending state of the last trellis step is equal to the starting state of the fist trellis step. However, this technique requires estimating the circulation state at the decoder side. Since training operations to estimate the circulation state would increase the SISO latency, an effective alternative [Zhan et al. 2006] is to inherit these metrics from the previous iteration.

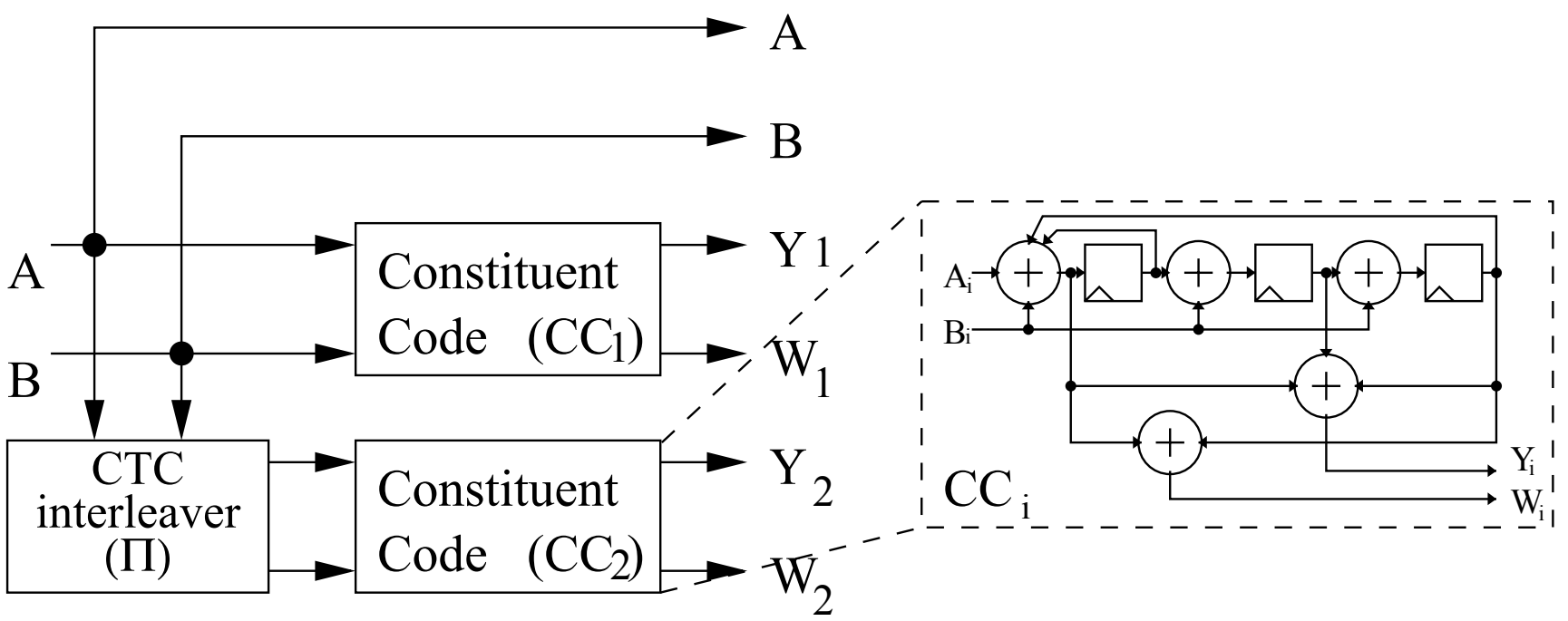

Fig. 6. WIMAX CTC: encoder and constituent CC structures

As in Viterbi decoder architectures often in CTC decoders the state metrics are computed by means of the "wrapping" representation technique proposed in [Hekstra, 1989]. This solution requires a normalization stage, depicted in Fig. 7, when combining $\alpha$, $\beta$ and $\gamma$ metrics to compute the extrinsic information as in (8). The last stage of the output processor, that computes the output extrinsic information, is a tree of max blocks for each component of the extrinsic information and few adders to implement (8). As highlighted in Fig. 7 this scheduling requires a buffer to store input LLRs that are used to compute the backward recursion (BMU-MEM). Since the output extrinsic information is computed during the backward recursion, forward recursion metrics are stored in a buffer (α-MEM). Further memory is required to implement the border metric inheritance, α-EXT-MEM, β-EXT-MEM and β-LOC-MEM.

The throughput sustained by the CTC decoder, defined as the number of decoded bits over the time required for their computation, is

$$T = \frac{k \cdot N_T \cdot f_{clk}}{2I \cdot (N_{cyc}^{SISO} + N_{cyc}^{ID})} = \frac{k \cdot N_T \cdot f_{clk}}{2I \cdot N_{cyc}^{dec}} \tag{15}$$

where $f_{clk}$ is the clock frequency, $N_T$ is the number of trellis steps, $k$=1 for a binary CTC, $k$=2 for a double binary CTC, $2I$ is the number of half iterations, $N_{cyc}^{SISO}$ and $N_{cyc}^{ID}$ represent the number of clock cycles required by one SISO and by the interleaving/deinterleaving structure. Since both $N_{cyc}^{SISO}$ and $N_{cyc}^{ID}$ are a function of $N_T$ they can be rewritten as $N_{cyc}^{SISO}=N_T \cdot SP+SISO_{cyc}^{lat}$ and $N_{cyc}^{ID}=N_T \cdot SP+ID_{cyc}^{oh}$ where $SP$ is the sending period, namely the rate sustained by the decoder to output two consecutive valid output data ($SP$=1 means at each clock cycle new valid output data are ready), $SISO_{cyc}^{lat}$ is the decoder latency, namely the number of clock cycles spent to produce the first valid output data, and $ID_{cyc}^{oh}$ is the interleaver/deinterleaver architecture overhead expressed in clock cycles. Usually, resorting to pipelining, $N_{cyc}^{SISO}$ and $N_{cyc}^{ID}$ can be partially overlapped; thus, the number of cycles required by one SISO decoder is $N_{cyc}^{dec}=N_T \cdot SP+SISO_{cyc}^{lat}+ID_{cyc}^{oh}$. Using the sliding window technique with the border metric inheritance strategy [Abbasfar & Yao 2003]; [Zhan et al. 2006] we obtain $SISO_{cyc}^{lat} \approx SP \cdot W$ and so (15) can be rewritten as (16), where the rightmost expression is obtained considering $W \ll N_T$ and $ID_{cyc}^{oh} \ll SP \cdot N_T$ that is a reasonable assumption in real cases.

$$T = \frac{k \cdot N_T \cdot f_{clk}}{2I \cdot [SP \cdot (N_T + W) + ID_{cyc}^{oh}]} \approx \frac{k \cdot f_{clk}}{2I \cdot SP} \tag{16}$$

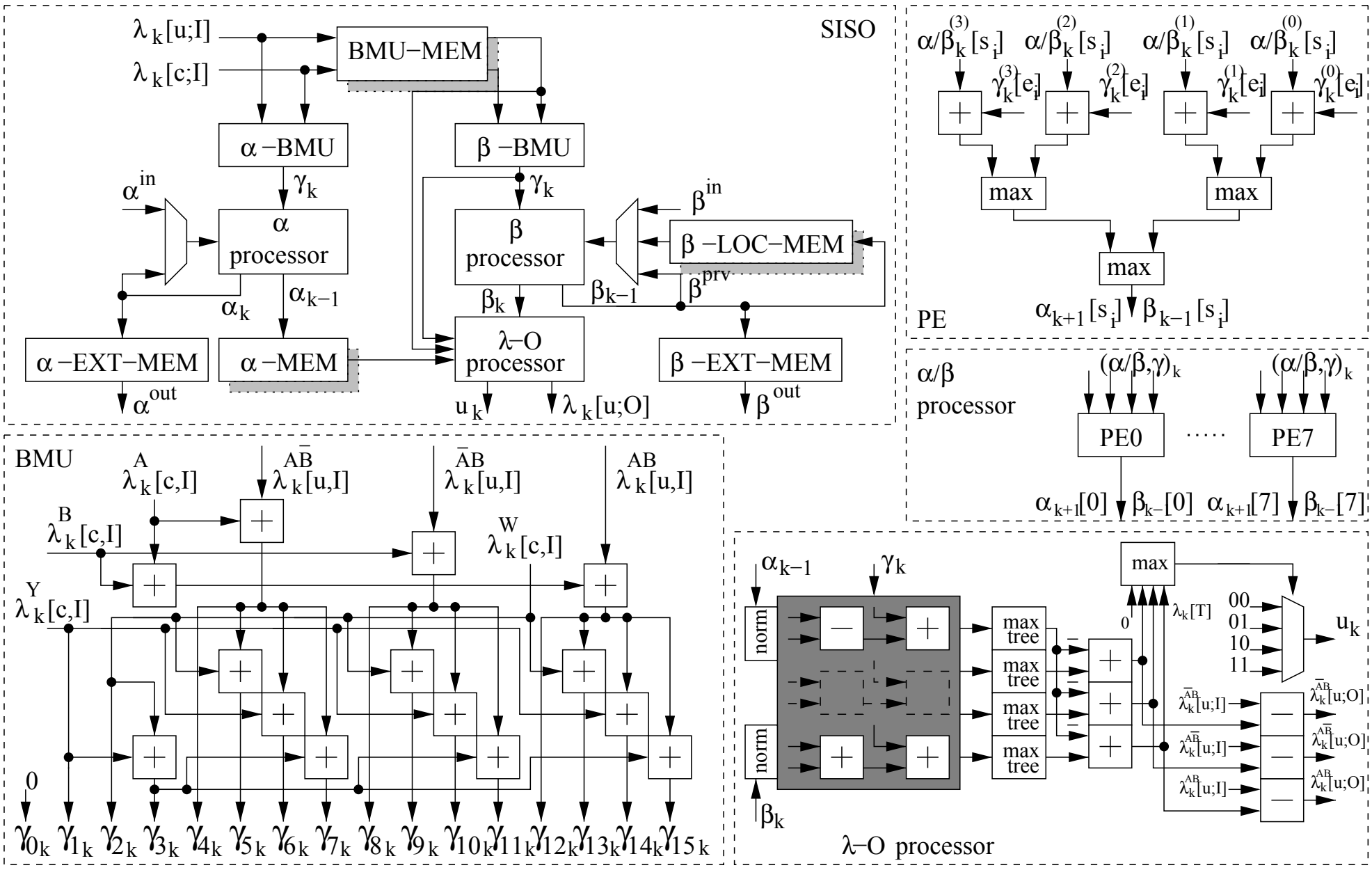

Fig. 7. WIMAX SISO block scheme

Usually optimized architectures [Masera et al., 1999]; [Bickerstaff et al., 2003]; [Kim & Park, 2008] are obtained with $SP$=1, whereas flexible architectures have higher $SP$ values [Vogt & Wehn, 2008]; [Muller et al., 2009]. However, even with $SP$=1, a double binary turbo decoder architecture that achieves the throughput imposed by WIMAX with eight iterations ($I$=8), would require $f_{clk}$=600 MHz. A possible solution to improve the throughput by a factor that ranges in [1.2, 1.9] is the based on decoder level parallelism [Muller et al. 2006] and is usually referred to as "shuffling" [Zhang & Fossorier, 2005]. However, to further improve the throughput a parallel decoder made of $P$ SISOs working concurrently is required. As a consequence, a parallel architecture achieves a throughput

$$T = \frac{k \cdot N_T \cdot f_{clk}}{2I \cdot [SP \cdot ({N_T}/{P} + W) + ID_{cyc}^{oh}]} \approx \frac{k \cdot P \cdot f_{clk}}{2I \cdot SP} \tag{17}$$

Thus, setting $P$=4, $I$=8 and $SP$=1, the WIMAX throughput is obtained with $f_{clk}$=150 MHz. It is worth pointing out that a $P$-parallel CTC decoder is made of $P$ SISOs connected to $P$ memories devoted to store the extrinsic information. However, in a parallel decoder during the scrambled half iteration collisions can occur, namely more SISOs could need to access the same memory during the same cycle. Since the collision phenomenon increases $ID_{cyc}^{oh}$, several algorithmic approaches to design collision free interleavers [Giulietti et al. 2002]; [Kwak & Lee, 2002]; [Gnaedig et al., 2003]; [Tarable et al., 2004] have been proposed. On the other hand, architectures to manage collisions in a parallel turbo decoder have also been proposed in the literature [Thul et al., 2002]; [Gilbert et al., 2003]; [Thul et al., 2003]; [Speziali & Zory, 2004]; [Martina et al. 2008-a]; [Martina et al., 2008-b], in particular [Martina et al. 2008-b] deals with the parallelization of the WIMAX CTC interleaver and avoids collision by the means of a throughput/parallelism scalable architecture that features $ID_{cyc}^{oh}=0$.
It is worth pointing out that parallel architectures increase not only the throughput but also the complexity of the decoder, so that some recent works aim at reducing the amount of memory required to implement SISO local buffers. In [Liu et al., 2007] and [Kim & Park, 2008] saturation of forward state metrics and quantization of border backward state metrics is proposed. Further studies have been performed to reduce the extrinsic information bit width by using adaptive quantization [Singh et al., 2008], pseudo-floating point representation [Park et al., 2008] and bit level representation [Kim & Park, 2009].

### 2.4 LDPC code decoders

LDPC codes were originally introduced in 1962 by Gallager [Gallager, 1962] and rediscovered in 1996 by MacKay and Neal [MacKay, 1996]. As turbo codes, they achieve near optimum error correction performance and are decoded by means of high complexity iterative algorithms.
An LDPC code is a linear block code defined by a $C \times B$ parity check matrix $H$, characterized by a low density of ones: $B$ is the number of bits in the code (block length), while $C$ is the number of parity checks. A one in a given cell of the $H$ matrix indicates that the bit corresponding to the cell column is used for the calculation of the parity check associated to the row. A popular description of an LDPC code is the bipartite (or Tanner) graph shown in Figure 8 for a small example, where $B$ variable nodes (VN) are connected to C check nodes (CN) through edges corresponding to the positions of the ones in $H$.
LDPC codes are usually decoded by means of an iterative algorithm variously known as sum-product, belief propagation or message passing, and reformulated in a version that processes logarithmic likelihood ratios instead of probabilities. In the first iteration, half variable nodes receive data from adjacent check nodes and from the channel and use them to obtain updated information sent to the check nodes; in the second half, check nodes take the updated information received from connected bit nodes and generate new messages to be sent back to variable nodes.
In message passing decoders, messages are exchanged along the edges of the Tanner graph, and computations are performed at the nodes. To avoid multiplications and divisions, the decoder usually works in the logarithmic domain.

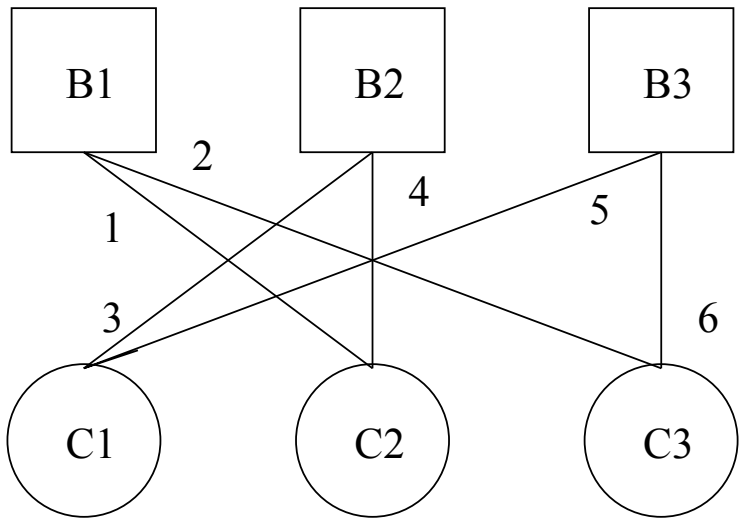

Fig. 8. Example Tanner graph

The message passing algorithm is described in the following equations, where $k$ represents the current iteration, $Q_{ji}$ is the message generated by VN $j$ and directed to CN $i$, $R_{ij}$ is the message computed by CN $i$ and sent to VN $j$. $C[j]$ is the whole set of incoming messages for VN $j$ and $R[i]$ is the whole set of the incoming messages for CN $i$.
Each variable node is initialized with the log-likelihood ratio (LLR) $\lambda_j$ associated to the received bit. Next, messages are propagated from the variable nodes to the check nodes along the edges of the Tanner graph. At the first iteration, only $\lambda_j$ are delivered, while starting from the second iteration VNs sum up all the messages $R_{ij}$ coming from CNs and combine them with $\lambda_j$ according to

$$Q_{ji}^{k} = \lambda_j + \sum_{\alpha \in C[j]/i} R_{\alpha j}^{k-1} \tag{18}$$

The check node computes new check to variable messages as

$$R_{ij}^{k} = \Psi^{-1}\left[ \sum_{\beta \in R[i]/j} \Psi\left[Q_{\beta i}^{k}\right] \right] \cdot \delta_{ij} \qquad \text{with} \qquad \delta_{ij} = (-1)^{|R[i]|} \prod_{j} \operatorname{sgn}(Q_{ji}) \tag{19}$$

where $|R[j]|$ is the cardinality of the CN and $\Psi(x)$ is a non linear function defined as

$$\Psi[x] = -\ln\left( \tanh\left|\frac{x}{2}\right| \right) \tag{20}$$

After a number of iterations that strongly depends on the addressed application and code rate (typically 5 to 40), variable nodes compute an overall estimation of the decoded bit in the form

$$\Lambda_j^{k} = \lambda_j + \sum_{\alpha \in C[j]} R_{\alpha j}^{k-1} \tag{21}$$

where the sign of $\Lambda_j$ can be understood as the hard decision on the decoded bit.

A large implementation complexity is associated to (19), which is simplified in different ways. First of all, function $\Psi(x)$ can be obtained by means of reduced complexity estimations [Masera et al., 2005]. Moreover sub-optimal, low complexity algorithms have been successfully proposed to simplify (19), such as for example the normalized Min-Sum algorithm [Chen et al., 2005] where only the two smallest magnitudes are used.
A further change is usually applied to the scheduling of variable and check nodes in order to improve communications performance. In the two-phase scheduling, the updating of variable and check nodes is accomplished in two separate phases. On the contrary, the turbo decoding message passing (TDMP) [Mansour & Shanbhag, 2003], also known as layered or shuffled decoding, allows for overlapped update operations: messages calculated by a subset of check nodes are immediately used to update variable nodes. This scheduling has been proved to be able to reduce the number of iterations by up to 50% at a fixed communications performance.
The required number of functional units in a decoder can be estimated based on the concept of processing power $P_c$ [Gouillod et al., 2007], which can be evaluated on the basis of the rate $R_c$ of the code, the number $K$ of information bits transmitted per codeword, the block size $N=K/R_c$, the required information throughput $D$, the operating clock frequency $f_{clk}$, the maximum number of iterations $i_{MAX}$ and the total number of edges to be processed per iteration $\varepsilon$. This relation is expressed as

$$P_c = \frac{\varepsilon \cdot D \cdot i_{MAX}}{K \cdot f_{clk}} \quad (22)$$

As two messages are associated with each edge (to be sent from the CN to the VN and vice versa), $2P_c$ gives the number of messages that must be concurrently processed at each decoding iteration in order to achieve the target throughput $D$. Equation (22) does not consider the message exchange overhead: yet it assumes that all messages dispatched during a cycle are delivered simultaneously during the same cycle. The $P_c$ value must then be assumed as a lower bound and the actual degree of parallelism strongly depends on both the structure of the $H$ matrix [Dinoi et al., 2006] and the adopted interconnect architecture among processing units [Quaglio et al., 2006] [Masera et al., 2007].
Actually, most of the implementation concerns come from the communication structure that must be allocated to support message passing from bit to check nodes and vice versa. Several hardware realizations that have been proposed in the literature are focused on how efficiently passing messages between the two types of processing units.
Three approaches can be followed in the high level organization of the decoder, coming to three kinds of architectures.

- Serial architectures: bit and check processors are allocated as single instances, each serving multiple nodes sequentially; messages are exchanged by means of a memory.
- Fully parallel architectures: processing units are allocated for each single bit and check node and all messages are passed in parallel on dedicated routes.
- Partially parallel architectures: more processing units work in parallel, serving all bit and check nodes within a number of cycles; suitable organization and hardware support is required to exchange messages.

For most codes and applications, the first approach results in slow implementations, while the second one has an excessive cost. As a result the only general viable solution is the third partially parallel approach, which on the other hand introduces the collision problem, already known in the implementation of parallel turbo decoders. Two main approaches have been proposed to deal with collisions:

- To design collision free codes [Mansour & Shanbhag , 2003], [Hocevar, 2003],
- To design decoder architecture able to avoid or at least mitigate collision effects [Kienle et al., 2003], [Tarable et al., 2004].

Even if the first approach has proven to be effective, it significantly limits the supported code classes. The second approach, on the other hand, is well suited for flexible and general architectures. An even more challenging task is the design of LDPC decoders that are flexible in terms of supported block sizes and code rates [Masera et al., 2007].

In partially parallel structures, permutation networks are used to establish the correct connections between functional units. However, structured LDPC codes, such as those specified in WIMAX, allow for replacing permutation networks by low complexity barrel shifters [Boutillon et al., 2000]; [Mansour & Shanbhag, 2003].

Early terminal schemes can be adopted to improve the decoding efficiency by dynamically adjusting the iteration number according to the SNR values. The simplest approach requires that decoding decisions are stored and compared across two consecutive iterations: if no changes are detected, the decoding is terminated, otherwise it is continued up to a maximum number of iterations. More sophisticated iteration control schemes are able to reduce the mean number of iterations, so saving both latency and energy [Kienle & When, 2005]; [Shin et al., 2007].

## 3. Case of study: complete WIMAX CTC decoder design

The WIMAX CTC decoder is made of three main blocks: symbol deselection (SD), subblock deinterleaver and CTC decoder as highlighted in Fig. 9 where $N$ represents the number of couples included in a data frame. SD, subblock deinterleaver and CTC decoder blocks are connected together by means of memory buffers in order to guarantee that the non iterative part of the decoder (namely SD and subblock deinterleaver) and the decoding loop work simultaneously on consecutive data frames. Since the maximum decoder throughput is about 75 Mb/s and the native CTC rate is 1/3 (two uncoded bits produce six coded bits), at the input of the decoding loop the maximum throughput can rise up to 225 millions of LLRs per second. The same throughput ought to be sustained by the subblock deinterleaver, whereas even higher throughput has to be sustained at the SD unit in case of repetition.

### 3.1 Symbol deselection

Depending on amount of data sent by the encoder (puncturing or repetition), the throughput sustained by the symbol deselection (SD) can rise up to 900 millions of LLRs per second (repetition ×4). When the encoder performs repetition, the same symbol is sent more than once. Thus, the decoder combines the LLRs referred to the same symbol to improve the reliability of that symbol. As shown in Fig. 9 this can be achieved partitioning the symbol deselection input buffer into four memories, each of which containing up to $6N$ LLRs.

Since the symbol deselection architecture can read up to four LLRs per clock cycle, it reduces the incoming throughput to 225 millions of LLRs per second. However, the symbol deselection has to compute the starting location and the number of LLRs to be written into the output buffer. The number of LLRs and the starting location are obtained as in (23) and (24) respectively, where $N_{SCHk}$, $m_k$ and $SPID_k$ are parameters specified by the WIMAX standard for the $k$-index subpacket when HARQ is enabled, namely $N_{SCHk}$, is the number of concatenated slots, $m_k$ is the modulation order and $SPID_k$ is the subpacket ID.

$$L_k = 48 \cdot m_k \cdot N_{SCHk} \tag{23}$$

$$F_k = (SPID_k \cdot L_k) \bmod 6N \tag{24}$$

Since $N_{SCHk} \in [1, 480]$ and $m_k \in \{2, 4, 6\}$ we can rewrite (23) as

$$L_k = \begin{cases} (2N_{SCHk} + N_{SCHk}) \cdot 2^5 & when \quad m_k = 2 \\ (2N_{SCHk} + N_{SCHk}) \cdot 2^6 & when \quad m_k = 4 \\ (8N_{SCHk} + N_{SCHk}) \cdot 2^5 & when \quad m_k = 6 \end{cases} \tag{25}$$

The efficient implementation of (25) is obtained with an adder whose inputs are $N_{SCHk}$ and the selection between two hardwired left shifted versions of $N_{SCHk}$ (one position and three positions), followed by a programmable left shifter (five-six positions). Similarly, since $SPID_k \in \{0, 1, 2, 3\}$, the multiplication in (24) is avoided as

$$F_k = \begin{cases} 0 & when \quad SPID_k = 0 \\ L_k \bmod 6N & when \quad SPID_k = 1 \\ 2L_k \bmod 6N & when \quad SPID_k = 2 \\ (2L_k + L_k) \bmod 6N & when \quad SPID_k = 3 \end{cases} \tag{26}$$

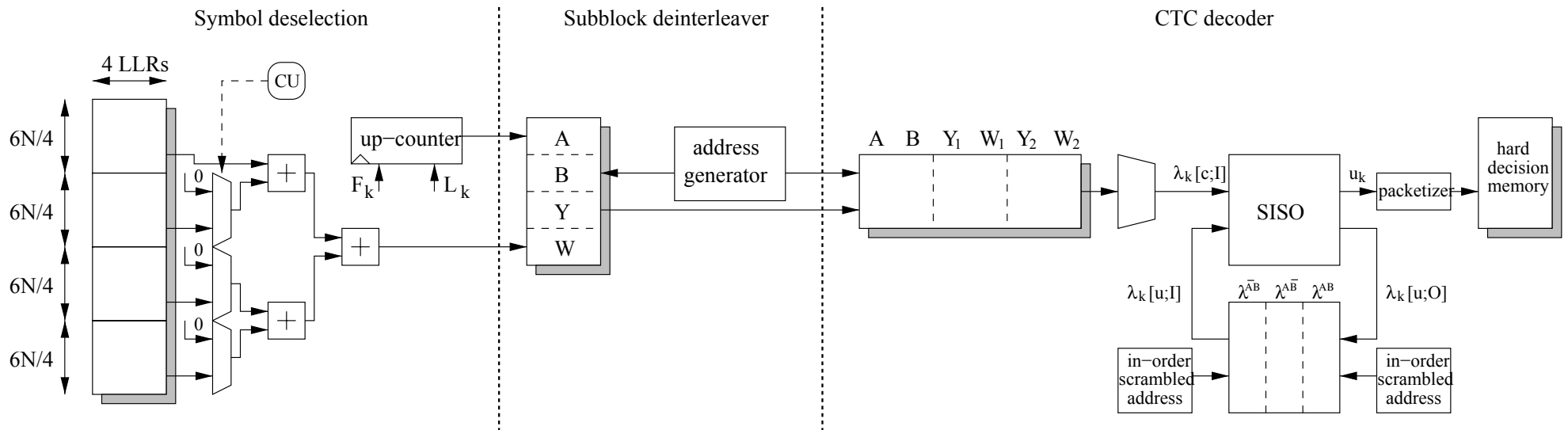

Fig. 9. Complete CTC decoder block scheme

A block scheme of the architecture employed to compute $F_k$ and $L_k$ is depicted in Fig. 10 (a). Furthermore, in order to support the puncturing mode, the output memory locations corresponding to unsent bits must be set to zero. To ease the SD architecture implementation, all the output memory locations are set to zero while $L_k$ and $F_k$ are computed. As a consequence, about two clock cycles per sample are required to complete the symbol deselection, namely 6N LLRs are output in 12N clock cycles. So that the symbol deselection throughput can be estimated as

$$T_{SD} = \frac{6N}{12N} f_{clk} = \frac{f_{clk}}{2} \tag{27}$$

As it can be observed, to sustain 225 millions of LLRs per second a clock frequency of 450 MHz is required. To overcome this problem we impose not only to partition the input buffer into four memories, but also to increase the memory parallelism, so that each memory location contains $p$ LLRs. Thus, we can rewrite (27) as (28) and by setting $p$ to a conservative value, as $p$=4, the SD architecture processes simultaneously up to sixteen LLRs with $f_{clk}$=113 MHz.

$$T_{SD} = \frac{6N}{12N/p} f_{clk} = \frac{p \cdot f_{clk}}{2} \tag{28}$$

### 3.2 Subblock deinterleaver

The received LLRs belong to six possible subblocks depending on the coded bits they are referred to ($A$, $B$, $Y_1$, $W_1$, $Y_2$, $W_2$) and each subblock is made of $N$ LLRs. The subblock deinterleaver treats each subblock separately and scrambles its LLRs according to Algorithm 1, given below, where $m$ and $J$ are constants specified by the WIMAX standard and $BRO_m(y)$ is the bit-reversed $m$-bit value of $y$.

$k \leftarrow 0$
$i \leftarrow 0$
**while** $i<N$ **do**
  $T_k \leftarrow 2^m(k \bmod J)+BRO_m(\lfloor k/J \rfloor)$
  **if** $T_k<N$ **then**
    $i \leftarrow i+1$
  **else**
    discard $T_k$
  **end if**
  $k \leftarrow k+1$
**end while**

Algorithm 1. Subblock deinterleaver address generator

As a consequence, the number of tentative addresses generated, $N_M$, can be greater than $N$. Exhaustive simulations, performed on the possible $N$ specified by the standard, show that the worst case is $N_M$=191 that occurs with $N$=144. Since 191/144=1.326, a conservative approximation is $N_M$=4$N$/3. The whole subblock deinterleaver architecture is obtained with one single address generator implementing Algorithm 1 to simultaneously write one LLR from each of the six subblock memories. In particular, as imposed by the WiMax standard, the interleaved LLRs belonging to the A and B subblocks are stored separately, whereas the interleaved LLRs belonging to $Y_1$ and $Y_2$ are stored as a symbol-by-symbol multiplexed sequence, creating a "macro-subblock" made of 2N LLRs. Similarly a macro-subblock made of 2N LLRs is generated storing a symbol-by-symbol multiplexed sequence of interleaved $W_1$ and $W_2$ subblocks.

Since all the subblocks can be processed simultaneously, this architecture deinterleaves six LLRs per clock cycle. As a consequence, the subblock deinterleaver sustains a throughput

$$T_{SubDein} = \frac{6N}{4N/3} f_{clk} = 4.5 f_{clk} \tag{29}$$

Thus, a throughput of 225 Millions of LLRs per second is sustained using $f_{clk}$=50 MHz.

To implement line 4 and 5 in Algorithm 1, three steps are required, namely the calculation of $k$ mod $J$ and $\lfloor k/J \rfloor$, the calculation of $2^m(k \bmod J)$ and $BRO_m(\lfloor k/J \rfloor)$, the generation of $T_k$ while checking $T_k<N$. It is worth pointing out that $k$ mod $J$ can be efficiently implemented as an up-counter followed by a mod $J$ block. Moreover, each time the mod $J$ block detects $k=J$, a second counter is incremented: the final value in the second counter is $\lfloor k/J \rfloor$. Since $m \in [3, 10]$, the $2^m(k \bmod J)$ term is implemented as a programmable shifter in the range [0, 7] followed by a hardwired three position left shifter. The $BRO_m(\lfloor k/J \rfloor)$ term is obtained by multiplexing eight hardwired bit reversal networks. Finally, a valid $T_k$ address is obtained with an adder and is validated by a comparator. The address generation architecture is shown in Fig. 10 (b).

### 3.3 CTC decoder

As detailed in section 2.3 to sustain the throughput required by the WIMAX standard a parallel decoder architecture is required. To that purpose we set $SP$=1, $I$=8, and $f_{clk}$=200 MHz, then from (17) we analyze the throughput as a function of $N$ for $W$=32. As shown in Fig. 11, only $P$=4 allows to achieve the target throughput (horizontal solid line) for $N \geq 480$.

Moreover, the window width impacts both on the decoder throughput and on the depth of SISO local buffers. So that a proper $W$ value for each frame size must be selected. In particular if $N/(P \cdot W) \in \mathbb{N}$ SISOs synchronization is simplified. However, the choice of $P$ should minimize collisions in memory access.

Exhaustive simulations show that collisions occur for $P$=2 and $P$=4 only with $N$=108. As a consequence, we select $P$ as a function of $N$ to simultaneously obtain a monotonically increasing throughput as a function of $N$ and to avoid collisions. It is worth pointing out that, when collisions are avoided, the resulting parallel interleaver is a circular shifting interleaver: the address generation is simplified with all SISOs simultaneously accessing the same location of different memories.

Said $idx^0_t$ the memory accessed by SISO-$0$ at time $t$ during a scrambled half iteration, the memory concurrently accessed by SISO-$k$ is $idx^k_t=(idx^0_t \pm k) \bmod P$.

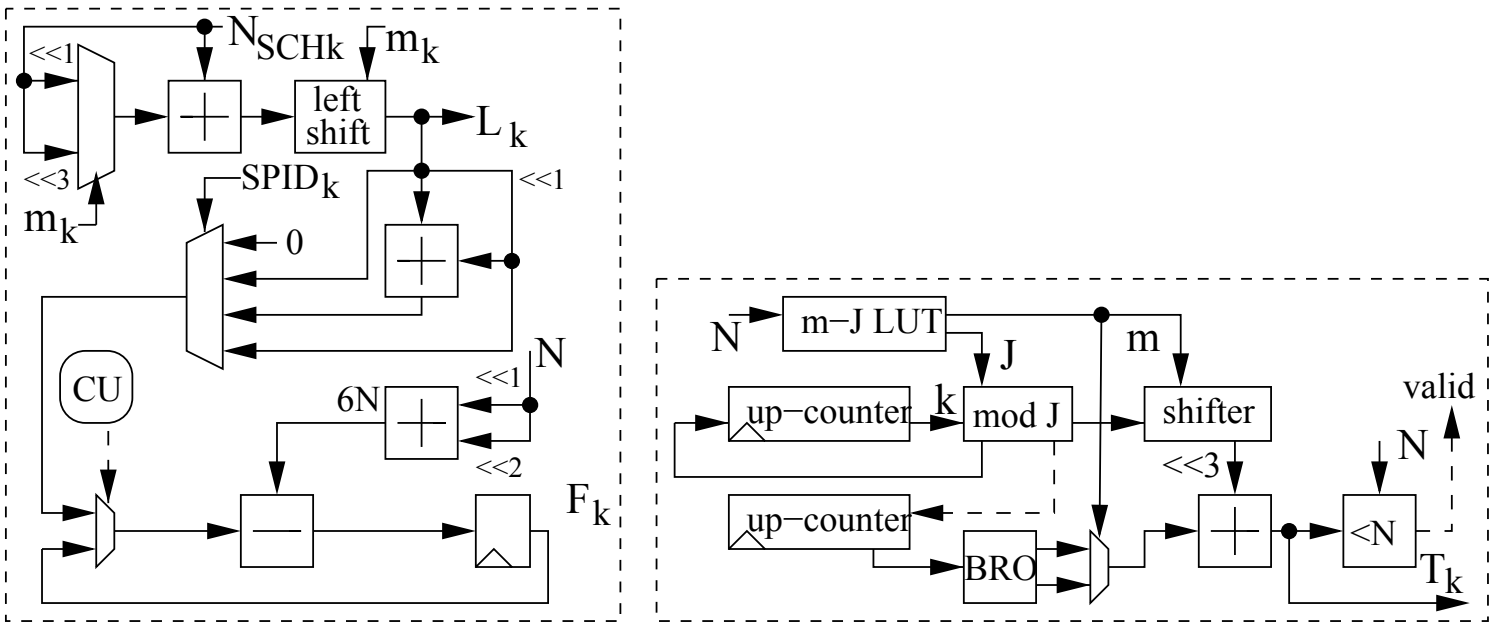

(a) (b)

Fig. 10. Symbol deselection starting address and number of elements generation block scheme (a), subblock deinterleaver address generation block scheme (b).

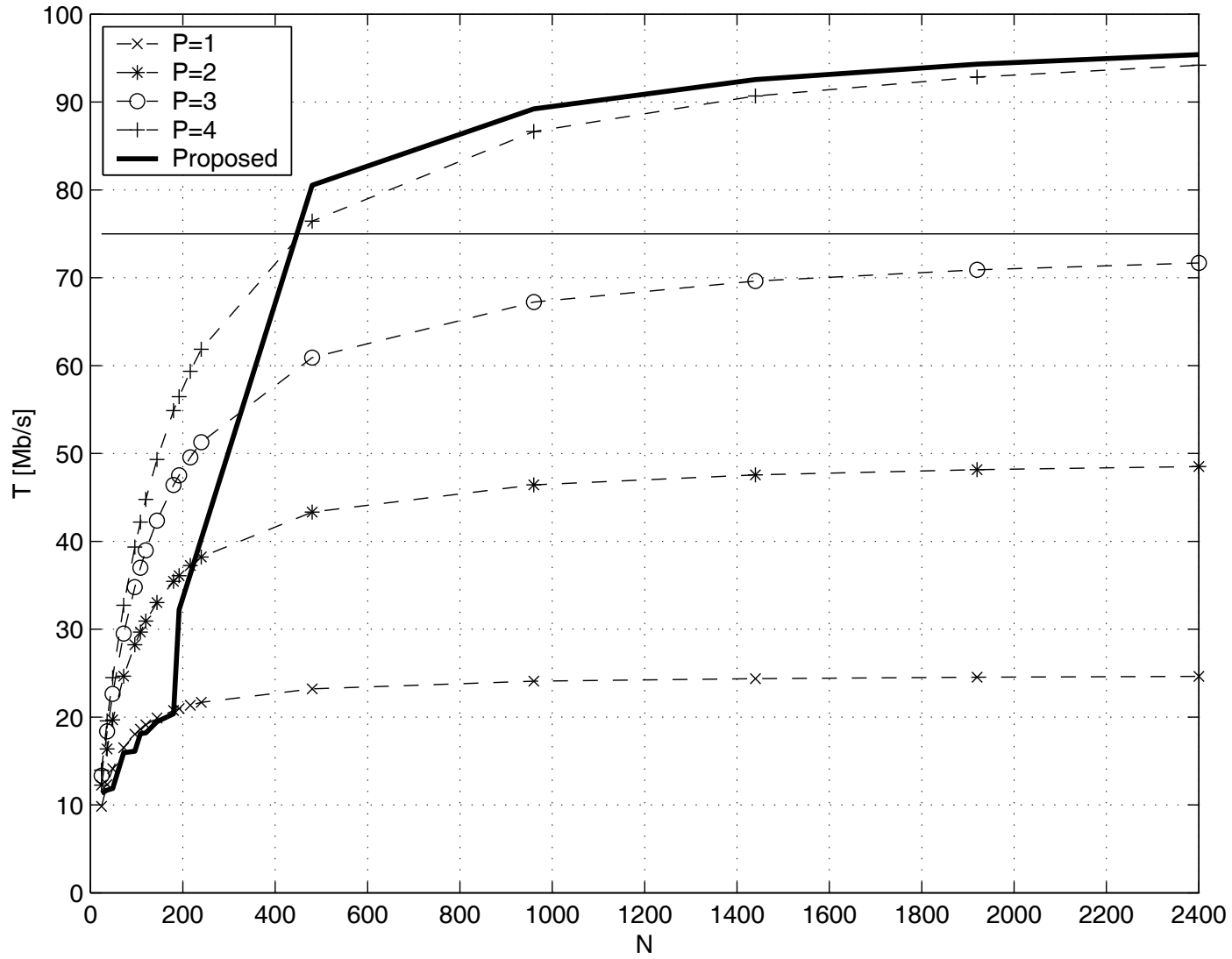

Fig. 11. Parallel CTC decoder throughput as a function of the block size N for different parallelism degree values P. The horizontal line represents the target throughput.

Thus, the parallel CTC interleaver-deinterleaver system is obtained as a cascaded two stage architecture (see Fig. 12). The first stage efficiently implements the WIMAX interleaver algorithm, whereas the second one extracts the common memory address $adx_t$ and the memory identifiers $idx^k{}_t$ from the scrambled address $i$.

The CTC interleaver algorithm specified in the WIMAX standard is structured in two steps. The first step switches the LLRs referred to $A$ and $B$ that are stored at odd addresses. The second step provides the interleaved address $i$ of the $j$-th couple as

$$i = (P_0 \cdot j + P_j^{'}) \bmod N \quad j = 0, \cdots, N-1 \tag{30}$$

where $P_0$ and $P_j'$ are constants that depend only on $N$ and are specified by the standard. It is worth pointing out that the two steps can be swapped, as a consequence the first step can be performed on-the-fly, avoiding the use of an intermediate buffer to store switched LLRs. A simple architecture to implement (30) can be derived by rewriting (30) as

$$i = \left[ i_j' + \left( P_j' \bmod N \right) \right] \bmod N \tag{31}$$

where

$$i_j' = \begin{cases} i_0' = 0 & when \quad j = 0 \\ \left( i_{j-1}' + P_0 \bmod N \right) \bmod N & when \quad j = 1, \cdots, N-1 \end{cases} \tag{32}$$

A small Look-Up-Table (LUT) is employed to store $P_0$ mod $N$ and $P_j'$ mod $N$ terms; then (31) is implemented by two parts as depicted in Fig. 12. The first part accumulates $P_0$ to implement the $P_0 \cdot j$ term and the mod $N$ block produces the correct modulo $N$ result. The second part employs the two least significant bits of a counter ($j-$cnt) to select the proper $P_j'$ mod $N$ value, which is added to the ($P_0 \cdot j$) mod $N$ term. A further modulo $N$ operation is performed at the output. Since in this architecture both the first and the second part work on data belonging to [0, $2N-1$], all the mod $N$ operations are implemented by means of a subtracter and a multiplexer.

The second stage of the parallel CTC interleaver-deinterleaver architecture works as follows. Since $adx_t \in [0, N/P-1]$, it can be obtained from the scrambled address $i$ produced by the first stage as

$$adx_t = \begin{cases} i & when & i \in \left[ 0, \frac{N}{P} - 1 \right] \\ i - \frac{N}{P} & when & i \in \left[ \frac{N}{P}, \frac{2N}{P} - 1 \right] \\ \cdots & \cdots & \cdots \\ i - (P-1)\frac{N}{P} & when & i \in \left[ (P-1)\frac{N}{P}, N-1 \right] \end{cases} \tag{33}$$

The straightforward implementation of (33) needs to calculate $N/P$ and to allocate $P-2$ multipliers, $P-1$ subtracters, a $P$-way multiplexer and few logic for selecting the proper $adx_t$ value. The $N/P$ division can be simplified by choosing the possible $P$ values as powers of two. Thus, we obtain a CTC decoder architecture that exploits throughput/parallelism scalability to avoid collisions, namely we employ: $P$=1 when $N \leq 180$, $P$=2 when $192 \leq N \leq 240$ and $P$=4 when $480 \leq N \leq 2400$. Moreover, as it can be inferred from Fig. 12, multiplications are avoided resorting to simple shift operations ($x >> i = x/2^i$). The sign of the subtractions (dashed lines in Fig. 12) allows not only to select the proper $adx_t$ but also to find $idx_t^0$. Then, with $P-1$ modulo $P$ adders the other $idx_t^k$ values are straightforwardly generated. As it can be observed, choosing $P$ as a power of two reduces the modulo $P$ adders to simpler, binary adders. The actual throughput sustained by the described throughput/parallelism scalable architecture is represented by the bold line in Fig. 11.

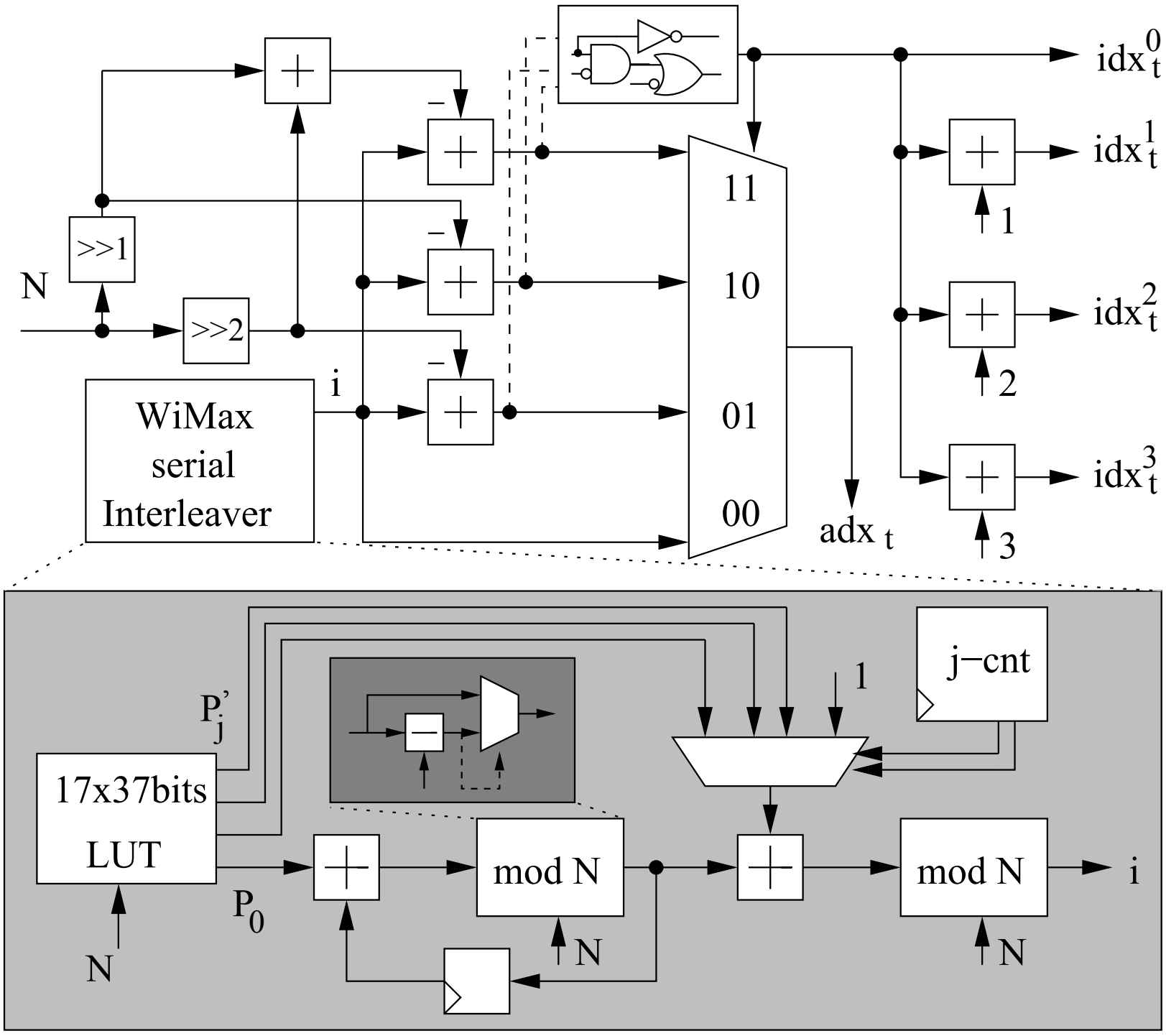

Fig. 12. Parallel CTC address generator

The global architecture of the designed parallel SISO is given in Fig. 13 where each SISO contains the processors devoted to compute the different metrics required by the BCJR algorithm as detailed in section 2.3. A simple network is used to properly connect the SISOs according to the current value of $P$ by setting the signal last_SISO. Furthermore, one address crossbar-switch (radx-switch) is used to implement the reading operation, a LIFO stores the address and makes them available for the writing phase, two data crossbar-switches (rdata-switch/wdata-switch) are used to properly send (receive) the data to (from) the memory (EI-MEM) according to the parallel interleaver $idx^k{}_t$ values.

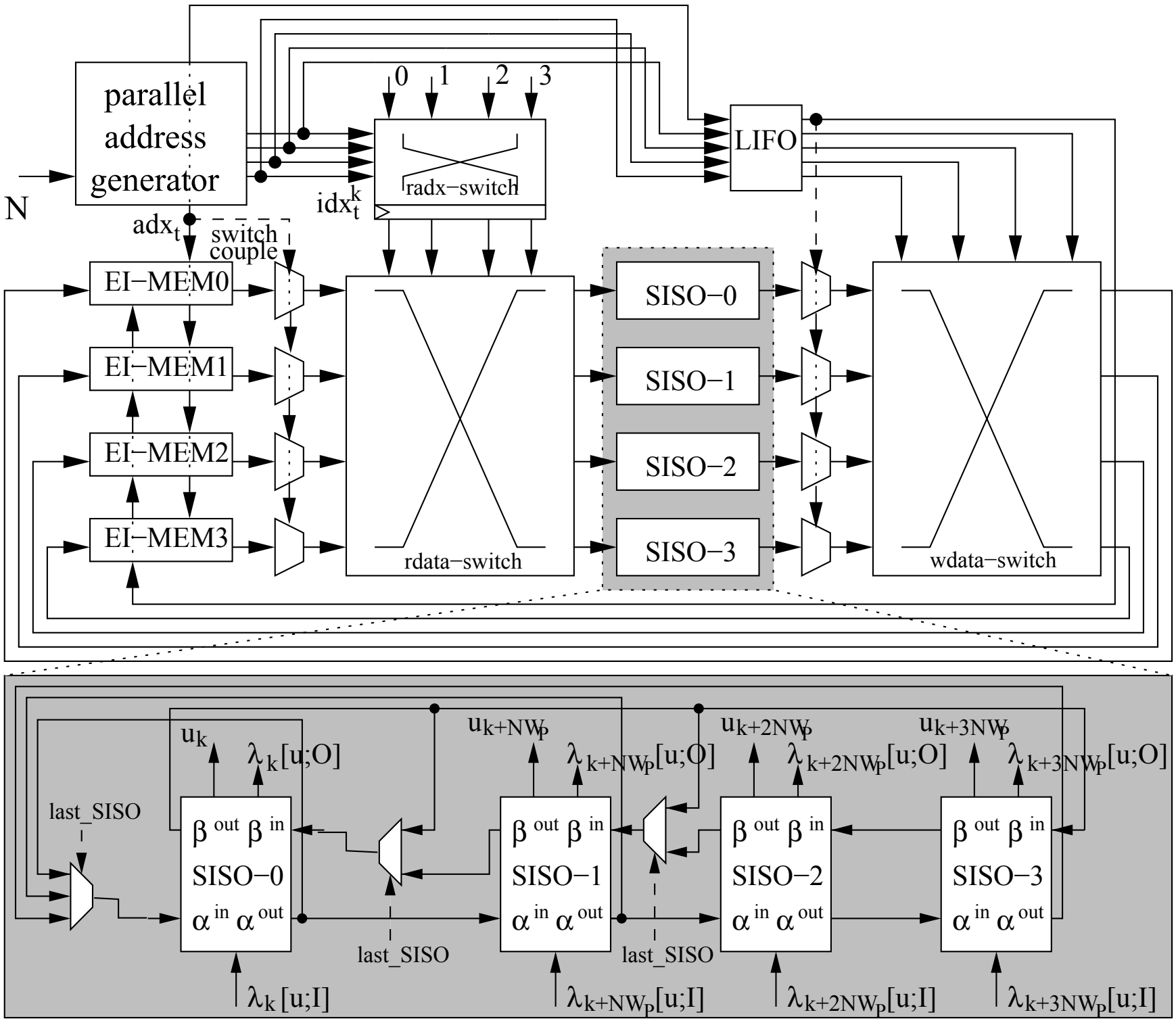

Fig. 13. Parallel CTC decoder architecture

In Table 2 the complexity of all the blocks for a 130 nm standard cell technology is reported. The bit-width is: 6 bit for $\lambda[c;I]$, 8 bit for $\lambda[u;I]$, and 12 bit for the state metrics. For further details the reader can refer to [Martina et al., 2009].

| *Architecture* | *SD* | *Subblock Deinterl.* | *SISOx1* | *Parallel Interl.* |
|---|---|---|---|---|
| Logic [kgate] | 11 | 1.7 | 37 | 2.8 |
| Memory [kbit] | 0 | 0 | 14.2 | 59 |

Table 2. Complexity of the whole receiver

## 4. Acknowledgements

This work is partially supported by the WIMAGIC project funded by the European Community.